\documentclass[a4paper,12pt]{article}
\usepackage{amssymb,amsmath,epsfig,esint}
\usepackage{CJKutf8}
\usepackage{amsfonts}
\usepackage{setspace,lipsum}
\usepackage{times}
\usepackage{multirow}
\usepackage{multicol}
\usepackage{setspace}
\usepackage[top=1.5cm,bottom=1.5cm,left=2cm,right=2cm,marginparwidth=1.5cm]{geometry}
\usepackage{mathptmx}
\usepackage[noadjust]{cite}    
\usepackage{calrsfs}
\DeclareMathAlphabet{\pazocal}{OMS}{zplm}{m}{n}

\newcommand{\Lb}{\pazocal{L}}

\usepackage{sidecap}
\sidecaptionvpos{figure}{t}

\usepackage[english]{babel}
\usepackage{booktabs}
\usepackage{tabu}
\usepackage[T1]{fontenc}
\usepackage{aecompl}
\usepackage{etoolbox} 

\usepackage[colorlinks=true, allcolors=blue]{hyperref}

\usepackage{graphicx}
\usepackage{ragged2e}
\usepackage{url}
\usepackage{xpatch}
\usepackage[printonlyused]{acronym}
\usepackage{xcolor}

\newcommand{\Planck}{\textit{Planck }}
\newcommand{\BASE}{\textit{BASE }}

\usepackage[explicit]{titlesec}
\titleformat{\section}
{\normalfont\normalsize\bfseries}{\thesection}{12pt}{\filcenter
\MakeUppercase{#1}} 
\titleformat{\subsection}
{\normalfont\normalsize\bfseries}{\thesubsection}{12pt}{\filcenter{#1}} 

\begin{document}

\title{\textbf{ Role of Spatial Curvature in a Dark Energy Interacting Model }}
\author{Trupti Patil\thanks{trupti19@iiserb.ac.in} \hspace{0.1cm} and \hspace{0.1cm}Sukanta Panda
\thanks{sukanta@iiserb.ac.in} \\
$^\ast$$^\dagger${\small Department of Physics, Indian Institute of Science Education and Research Bhopal,}\\
{\small Bhopal - 462066, India}\\
}
\date{}
\maketitle
\begin{abstract}
    This paper investigates the effects of spatial curvature in a model where dark matter and dark energy interact. The analysis employs a range of datasets, including CMB, BAO, Type Ia Supernova, $H (z) $ from cosmic chronometers, $H_0$ measurements from Megamasers and SH0ES, growth rate data and strong lensing time delay measurements, to assess the model’s fit and explore the late-time dynamics of the interacting dark sector in a non-flat cosmological framework. The study indicates that introducing curvature does affect the Hubble constant ($H_0$) and the structure growth parameter ($S_8$), and also helps in alleviating the tensions between early and late universe observations to some extent. The observational data shows an indication for an open universe. This implies that the presence of curvature and its influence on Universe's evolution cannot be neglected entirely. 
\end{abstract}

\section{Introduction}
Understanding the intricate balance of energy forms throughout the vast expanse of the universe, as well as characterizing their perturbations, are primary goals of both ongoing and forthcoming cosmological observations of the Cosmic Microwave Background (CMB) and the Large-Scale Structure (LSS) of the cosmos. A robustly supported perspective in cosmology is the adoption of the standard flat $\Lambda$CDM cosmological model, which is backed by extensive observational data. This is often based on the findings of numerous studies that use data from sources such as the Cosmic Microwave Background (CMB), Baryon Acoustic Oscillations (BAOs) and Type Ia supernovae (SNe-Ia) to constrain various cosmological parameters, including the curvature density parameter $\Omega_k$ (see, for instance, \footnote{In Ref. 1, the authors conclude |$\Omega_k$|<0.0094 (95$\%$ confidence) by combining data from the WMAP experiment and other sources. They remark that ``a small deviation from flatness is expected and is worthy of future searches.''}Refs. \cite{Hinshaw_2013,Balbi:2000tg,MacTavish:2005yk,Melchiorri_2000,Planck:2015fie,SDSS:2003eyi,SDSS:2005xqv,SDSS:2006lmn,Dodelson:1999am,Wang:2001gy,Seljak:2006bg,Tegmark:2000qy}). The majority of these studies have shown that a flat universe is consistent with cosmological observations. However, recent observational probes \cite{Handley:2019tkm,DiValentino:2019qzk,Zhang:2021djh,Arjona:2021hmg,Gonzalez:2021ojp,Jesus:2019jvk,Luciano:2022ffn,Wang:2022rvf,Qi:2022sxm,Amendola:2024gkz} that attract considerable attention for the possibility of non-zero $\Omega_k$ values, which can influence the cosmic energy budget, add intrigue to this complex landscape. Also, there are well-studied curvature models motivated by string theory constructions, recently proposed in \cite{Andriot:2023wvg, Andriot:2024jsh} and investigated through observations in \cite{Bhattacharya:2024hep}. In this study, we further explore the possibility of non-zero spatial curvature.

\textit{Planck} CMB power spectra, when analysed with the \textit{Plik} likelihood, prefers a closed Universe $\Omega_k$ < 0 at $\sim$ 3$\sigma$ level \big($\Omega_k = -0.044^{+0.018}_{-0.015}$\big). Whereas the joint constraint with BAO measurements shows a preference for a flat universe with $\Omega_k = -0.001\pm 0.002$ \cite{Planck:2018vyg}, indicating a discrepancy in $\Omega_k$, between CMB and BAO measurements. When the \Planck temperature and polarization data are combined with the lensing reconstruction,  yield fits consistent with flatness to within $\sim$ 1$\%$  ($\Omega_k = -0.0106\pm 0.0065$). These constraints with BAO data addition, yield $\Omega_k = -0.0007\pm 0.0019$ \cite{10.1093/mnras/stx721,Planck:2018vyg},  corresponding to a 1$\sigma$ detection of flatness at an accuracy of 0.2$\%$. As an independent probe of cosmic curvature, a recent BAO compilation from the recent extended Baryon Oscillation Spectroscopic Survey (eBOSS) \cite{eBOSS:2020yzd} found $\Omega_k$ to be $0.078^{+0.086}_{-0.099}$. Additionally, an indication for a closed spatial curvature is also present in BAO data, using Effective Field Theories of LSS (EFTofLSS) \cite{Glanville:2022xes}, once the assumptions of flatness are removed from the beginning. These inconsistencies in the data for $\Omega_k$ constraints themselves give rise to what can be referred to as the curvature tension, as mentioned in \cite{Handley:2019tkm}. Also, in \cite{DiValentino:2019qzk,Handley:2019tkm,Vagnozzi:2020rcz,Qi:2023oxv}, they provide evidence for the possible discord among early and late universe observational probes (including BAO, Supernovae, lensing, and cosmic shear measurements) in $\Lambda$CDM models which permit curvature. These findings challenged the spatial flatness assumptions present in the standard $\Lambda$CDM model and sparked a debate about the possibility of a small but non-zero spatial curvature in our universe. Notably, observations consistent with near-flat geometry cannot conclusively establish that the universe is spatially flat \cite{Anselmi:2022uvj} unless supported by strong theoretical reasoning. Furthermore, incorrectly assuming a flat geometry for a spatially curved universe could significantly impact the inferred values of cosmological parameters \cite{Dossett:2012kd}. Since allowing for spatial curvature can impact parameter constraints \cite{Dossett:2012kd}, it is essential to consider the curvature parameter and allow the observational data to determine the most likely scenario, avoiding possible biases introduced by the flatness assumption.




Furthermore, the authors of \cite{Yang:2022kho,DiValentino:2019qzk,DiValentino:2020hov,deCruzPerez:2024shj,Yang:2021hxg} have found that the spatial curvature may significantly influence tension in the Hubble expansion parameter, $H_0$, known as the Hubble tension, as well as the tension in the structure growth parameter, $S_8$, referred to as the $S_8$ tension. In \cite{deCruzPerez:2024shj,Benetti:2021div,DESI:2024mwx,Liu:2022mpj} it is shown that though the impact of curvature is small, it does alter the $H_0$ and $S_8$ constraints when compared to results assuming flat spatial geometry. In \cite{Bolejko:2017fos}, the authors presented that emerging spatial curvature carries the potential to resolve the Hubble constant tension\footnote{In the appendix, we provide the physical explanation of why and how spatial curvature can help resolve the Hubble tension.} between high-redshift (CMB) and low-redshift (distance ladder) measurements, as shown using generated mock supernovae data. Therefore, in this manuscript, we aim to investigate this possibility of additional spatial curvature in the universe and its implications on $H_0$ and $S_8$ tensions. In particular, we intend to study the impact on the value of the Hubble constant and the existing more than 5$\sigma$ tension between \Planck CMB(2018) + $\Lambda$CDM estimate \cite{Planck:2018vyg} and the SH0ES \cite{Riess:2021jrx} measurement, together with the $\sim$3$\sigma$ tension of \Planck CMB data with weak lensing measurements and redshift surveys \cite{Heymans:2020gsg,DES:2021vln,DES:2022ccp}, in the value of $S_8$ \cite{DIVALENTINO2021102604}. The tension can be readily visualized as the combination of the amplitude of matter clustering $\sigma_8$ in the late universe and the present matter density, $\Omega_{m_0}$, given by $S_8$ = $\sigma_8 \sqrt{\Omega_{m_0}/0.3}$. Several works \cite{Yang:2018euj1,Wang_2016,Adil:2023exv,Verde:2019ivm,Evslin:2017qdn,Yang:2018uae,Karwal:2021vpk,Freese:2021rjq,Cruz:2022oqk,FrancoAbellan:2021sxk,Okamatsu:2021jil,Banerjee:2020xcn,Lee:2022cyh,Chowdhury:2023opo,Montani:2023xpd,deAraujo:2021cnd} tried to address these cosmological tensions but could only resolve one tension or the other while maintaining the same number of parameters. The only solutions proposed by \cite{Knox:2019rjx,DiValentino:2020zio, DiValentino:2021izs,Sen:2021wld, Escamilla:2023shf,Adil:2023exv,mehdi22,mehdi21,Patil:2022ejk,Patil:2022uco,mehdi19}  to resolve all cosmological tensions at present are either highly fine-tuned or involve introducing additional parameters, thereby increasing uncertainties in the parameter space. 

Among various proposed solutions, interacting dark sector models \cite{Barros:2018efl,PhysRevD.62.043511,Abdalla:2022yfr,Patil:2023rqy,Lucca:2020zjb,Gao:2022ahg,Kumar:2017dnp,Gariazzo:2021qtg,Salvatelli:2014zta,Richarte:2015maa,Nunes:2016dlj,DiValentino:2020kpf}, stand out as potential avenues for addressing tensions related to the $H_0$ and the $S_8$ parameter in alternative cosmological frameworks. Motivated by this interesting possibility, we investigate whether interacting dark sector scenarios can mitigate the discrepancies between early- and late-time observational measurements within non-flat cosmology. In this article, while similar aspects have been explored in the past \cite{Patil:2023rqy}, we present two new insights within the dark sector. First, we introduce an additional parameter of spatial curvature within the framework of dark sector interactions that may effectively alleviate or resolve these tensions. Second, we utilize new observational data, including eBOSS DR16 BAO data \cite{eBOSS:2020fvk,eBOSS:2020abk,eBOSS:2020lta,eBOSS:2020gbb} and time-delay strong gravitational lensing measurements from H0LiCOW experiments \cite{Bonvin:2016crt}. 

The work is organized as follows. The following section explains the model framework, the basic equations and the linear dark matter perturbation evolution equation for the coupled dark sector assuming a non-flat background of the universe. Section \ref{observational analysis} describes the observational data and methodology employed in our data analysis pipeline. We present the results obtained in our study in Section \ref{results and discussion}, and finally, in Section \ref{conclusions}, we conclude the findings.

\section{Interacting Dark Energy in a Curved Universe}\label{model dynamics}
Assuming the large-scale homogeneous and isotropic space-time, the geometry of the universe is characterized by the Friedmann-Lema$\hat{i}$tre-Robertson-Walker (FLRW) line element as given by
\label{eqn:line element}
\begin{equation}
    \text{d}s^2 = -c^2\text{d}t^2 + a^2(t) \bigg[\frac{\text{d}r^2}{1-Kr^2}+r^2\text{d}\theta^2+r^2 \sin^2 \theta \text{d}\phi^2 \bigg].
\end{equation}
where `$c$' is the light speed, `$a(t)$' is the scale factor as a function of cosmic time, and `$K$' is the curvature parameter. The values $K$ = 0, +1, -1, respectively, denote the universe's spatially flat, closed, and open geometry. We consider the dark energy as a canonical scalar field $\phi$, interacting with cold dark matter element ($\omega_{dm}$=0). Following this assumption, the action for the dark sector interaction can be written as 
\begin{equation}
\label{eqn:action integral}
    S =  \int \text{d}^4x \sqrt{-g}\bigg[ \frac{R}{2}  - \frac{1}{2}  g^{\mu \nu}  \partial_{\mu}\phi \partial_{\nu}\phi - V(\phi) + \sum_{i} {\Lb}^{i}_m (\chi_i, \phi) \bigg].
\end{equation}
$\Lb_m$ is the matter Lagrangian for different matter field elements, which shows $\phi$ dependency through the coupling. Different matter species ($i$) may experience different couplings \cite{PhysRevD.62.043511,Amendola:2001rc}. Our study considers a cosmological scenario where only cold dark matter (CDM) couples to the dark energy (DE) scalar field $\phi$ by energy or momentum exchange mechanism between them in a non-gravitational way. The other fluids, such as radiation and baryons, are assumed uncoupled to dark energy \cite{PhysRevD.43.3873}. The previous study \cite{Patil:2023rqy} analysed the action integral (Eq. (\ref{eqn:action integral})) for $K$ = 0 case. Unlike the earlier analysis, here, we perform the analysis for non-zero curvature case.\\
The variation of Eq. (\ref{eqn:action integral}) w.r.t. the inverse metric gives Einstein’s gravitational field equation as
\begin{equation}
\label{eqn:3}
    G_{\mu \nu} = T_{\mu\nu} \equiv T_{\mu \nu} ^{(\phi)} + T_{\mu \nu} ^{(m)}.
\end{equation}
where $T_{\mu \nu}$ is the sum of $T_{\mu \nu}^{(\phi)}$ and $T_{\mu \nu}^{(m)}$ which are energy-momentum tensor of DE component and matter component, respectively. By permitting interaction between the dark species, the local conservation equation (Eq. (\ref{eqn:conservation eq})) for CDM and the DE scalar field at the background level become intertwined through a coupling or interaction function $Q$:
\begin{equation}
\label{eqn:conservation eq}
    -\nabla^\mu T_{\mu \nu} ^{(\phi)} = Q_{\nu}=  \nabla^\mu T_{\mu \nu} ^{(dm)},
\end{equation}
where $Q_{\nu}$ expresses the interaction between dark matter and dark energy 
\begin{equation}
\label{eqn:interaction strength}
    Q_{\nu} = F_{,\phi} \rho_{dm} \nabla_{\nu}{\phi}, 
\end{equation}
with $F_{,\phi} \equiv \partial F / \partial \phi$. $F(\phi)$ = $F_0$ e$^{\beta \phi}$ is the coupling strength and $\beta$ is a constant. As the radiation component does not interact with dark species, it conserves independently and follows $\nabla^\mu T_{\mu \nu} ^{(r)} = 0$.
Utilizing the above equations, we obtain the two Friedmann equations for the background evolution as
\begin{equation}
\label{eqn:background eq}
\begin{split}
\frac{\dot{\phi}^2}{2}+ V(\phi) + F(\phi) \rho_{dm} +\rho_{r} - \frac{3K}{a^2}= 3H^2, \\[4pt]
-\bigg[\frac{\dot{\phi}^2}{2}- V(\phi) + F(\phi) P_{dm} +P_{r}\bigg] - \frac{K}{a^2}= 2\dot{H} + 3H^2.
\end{split}
\end{equation}
and the continuity equation for each component as
\begin{equation}
\label{eqn:continuity eq}
\begin{split}
\dot {\rho_{\phi}} +3H \rho_{\phi} (1+\omega_{\phi}) = Q,\\
\dot\rho_c +3H \rho_c  = -Q,\\
\dot {\rho_{r}} +3H \rho_{r} (1+\omega_{r}) = 0.
\end{split}
\end{equation}
\vspace{0.1cm}
where $H \equiv \frac{\dot{a}}{a}$, represents the time evolution of the universe. The dot indicates derivative w.r.t. time $t$, and we re-scale the term $``F(\phi)\rho_{dm}$'' as $\rho_c$, denoting the energy density for dark matter coupled to dark energy. 
Corresponding time part of interaction term in Eq. {\eqref{eqn:interaction strength}} corresponds to $Q  = F_{,\phi} \rho_{dm} \dot{\phi}$. Here, the interaction function $Q_\nu$ is derived systematically from the action integral and conservation equations, ensuring a robust correlation between dark matter and dark energy. This contrasts with phenomenological fluid models, where the interaction term is introduced ad hoc, without a formal derivation.


\vspace{0.1cm}
The coupling also modifies the matter perturbation evolution equation. In the linearized approximation, assuming the Newtonian gauge, the equation describing the growth of DM density perturbation in the sub-Hubble limit can be expressed as   \\
%
%
\begin{equation}
\label{coupled perturbation eq1}
        \delta_{c}^{''} +  \delta_{c}^{'} \bigg[2+\frac{H^{'}}{H} - \frac{F,_{\phi}}{F(\phi)} \phi^{'} \bigg] - \frac{3}{2} \delta_{c} \bigg[ \frac{{\rho_c }}{3H^2} - \frac{1}{3} \phi^{'2} \bigg] \big[1+  \frac{2 {F,_{\phi} }^2}{F(\phi)^{2}} \big]  =0.
\end{equation}
where, $\delta_c \equiv \frac{\delta \rho_{c}}{\rho_{c}}$ represents the evolution of matter density contrast and the subscript $c$ stands for the CDM component. $\delta \rho_{c}$ is the perturbed dark matter density. `$\prime$' expresses the derivatives w.r.t. the e-fold, $N$ = $ln(a)$. The governing equations, both at the background and perturbation levels, determine the dynamics of the interacting scalar field.

Finally, to study the dynamics of the system, we simplify the analysis by formulating autonomous equations  through the definition of dimensionless variables as follows:
\begin{align}
\label{dimensionless variables}
x &= \frac{{\dot \phi}}{{\sqrt{6} H}}, &
y &= \frac{{\sqrt{V(\phi)/3}}}{{H}}, &
\Omega_{dm} &= \frac{{F(\phi) \rho_{dm} }}{{3H^2}}, &
\Omega_{r} &= \frac{{\rho_{r}}}{{3H^2}}, &
\Omega_K &= \frac{{-K}}{{a^2 H^2}}, &
\end{align}
satisfying the constraint equation:
\begin{equation}
\label{constraint eq}
\Omega_{\phi} + \Omega_{dm} + \Omega_{r} + \Omega_K = 1.
\end{equation}
Here, $\Omega_{\phi} = x^2 + y^2$ represents the density parameter of dark energy. $\Omega_{dm}$ and $\Omega_{r}$ represent the density parameters of dark matter and radiation, respectively. Additionally, we define:
\begin{equation}
\label{exta variables}
\lambda = -\frac{{V,{\phi}}}{{V(\phi)}}, \quad m = \frac{{F,{\phi} }}{{F(\phi)}}.
\end{equation}
These definitions contribute to forming a complete autonomous system.
We introduce an additional dimensionless variable `$\gamma$' to characterize the dark energy equation of state `$\omega_{\phi}$' as
\begin{equation}
\label{gamma def}
    1+\omega_{\phi} =\frac{2x^2}{x^2+y^2} = \gamma.
\end{equation}
Operating derivative with respect to number of e-folding $N = ln(a)$ on above variables, Eqs. \eqref{dimensionless variables}, \eqref{exta variables}, and \eqref{gamma def}, we can obtain the coupled dynamical set of equations within curved cosmology. During the dynamical analysis, we assumed an exponential potential, $V(\phi) \propto e^{-\lambda \phi}$ (where $\lambda$ is a constant), which results in the equation corresponding to $\lambda$ simplifying to $\lambda^\prime = 0$. Similarly, for the coupling term $F(\phi) \propto e^{\beta \phi}$, $m^\prime = 0$. These simplifications lead to the 4D system describing background dynamics, which can be expressed as
%
\begin{equation}
\label{autonomous system}
\begin{split}
\Omega_{\phi}^\prime =  3(1-\gamma)\Omega_{\phi} (1-\Omega_{\phi})  + \Omega_{\phi} \Omega_{r} + m \Omega_{dm} \sqrt{3\gamma \Omega_{\phi}}  - \Omega_{\phi} \Omega_K ,\\[0.5pt]
\Omega_{r}^\prime =  \Omega_{r}(\Omega_{r} - 1) + 3 \Omega_{r} \Omega_{\phi} (\gamma - 1) - \Omega_{r} \Omega_K  ,\\[0.5pt]
\Omega_K^\prime = - \Omega_K (\Omega_K - 1) + 3 \Omega_K \Omega_{\phi} (\gamma - 1)  + \Omega_{r} \Omega_K, \\[0.5pt]
\gamma^\prime = \big(2-\gamma \big) \sqrt{3\gamma \Omega_{\phi}}  \bigg(-\sqrt{3\gamma \Omega_{\phi}} +\lambda \Omega_{\phi} +  m  \Omega_{dm} \bigg).  \\[0.5pt]
\end{split}
\end{equation}
%
%
%
%
The initial condition on $\omega_\phi$ is $\omega_\phi \approx -1$, i.e., $\gamma \simeq 0.0001$ \cite{Caldwell:2005tm}. Other parameters, namely $\Omega_\phi$, $\Omega_r$, $\Omega_K$, $\lambda$, and $m$, are kept as free parameters.

DM density perturbation equation Eq. \eqref{coupled perturbation eq1} describing the perturbed dynamics can now be written as
%
\begin{equation}
\label{coupled perturbation eq2}
        \delta_{c}^{''} = - \bigg(\frac{1}{2} -\frac{3}{2} \Omega_{\phi}(\gamma-1) - \frac{1}{2} \Omega_{r} +\frac{1}{2} \Omega_K   - m \sqrt{3\gamma \Omega_{\phi}} \bigg) \delta_{c}^{'} + \frac{3}{2}\bigg( 1-\Omega_{\phi}(1+\gamma)-\Omega_{r} - \Omega_K \bigg) \big(1+ 2m^2 \big)    \delta_{c} .
\end{equation}
To ensure the universe's evolution around the matter-dominated epoch, we assume the time evolution at $N_i$ = $-7$. Hence, we put the initial conditions as $\phi(N_i)$ = $\phi^{\prime}(N_i)$ = 0 and $\delta_{c}(N_i)$ = $\delta_{c}^{'}(N_i)$ = $10^{-3}$. Combining Eq. \eqref{autonomous system} and Eq. \eqref{coupled perturbation eq2}, background and perturbed set of equations, we constrain the model and its parameters using the cosmological observations as described in the next section.

\section{Observational Data and Methodology}\label{observational analysis}
In this section, we present the latest observational data from diverse sources used to investigate and constrain model parameters in the case of interactions with curvature.
\begin{itemize}
    \item \textbf{CMB:} In the present analysis, we consider the CMB distance priors from the final \Planck CMB data obtained by Chen et al \cite{Chen:2018dbv}. In particular, we have used \textit{Planck} 2018 compressed CMB likelihood TT, TE, EE + lowE \cite{Planck:2015bue}.

    Note that we employ CMB distance priors for the 
    efficient and self-consistent approach they offer in exploring the parameter space of our model. The distance priors have been effectively used to study the late-time universe and implemented to replace global fitting of the full Planck 2018 data for dark energy models beyond standard $\Lambda$CDM \cite{Zhai:2019nad,Rezaei:2019xwo,Arjona:2019rfn,Zhai:2018vmm,Chen:2018dbv,Lazkoz:2006gp,Corasaniti:2007rf}. However, it is important to note that these priors may not fully capture the complexities of the full CMB power spectrum due to hidden assumptions, particularly in cases of significant deviations from the $\Lambda$CDM model, which may lead to systematic offsets in cosmological parameters \cite{Corasaniti:2007rf,Elgaroy:2007bv}. This assumption underlies our analysis. Thus, we clarify that the constraints derived from our work are based on the premise that the distance priors adequately represent the relevant features of the full CMB data in this case.

    \item \textbf{BAO:} For BAO distance measurements, we use data points from different experiments that include isotropic BAO measurements at small redshifts from 6dF \cite{Beutler_2011} and SDSS DR7-MGS surveys \cite{Ross:2014qpa}, as well as at high redshifts from SDSS DR14-eBOSS quasar clustering \cite{Ata:2017dya} and cross-correlation of Ly$\alpha$ with quasars from SDSS DR12 \cite{duMasdesBourboux:2017mrl}. Additionally, we consider anisotropic BAO measurements by BOSS DR12 galaxy sample \cite{BOSS:2016wmc}. We also added eBOSS DR16 data on emission line galaxies (ELG), luminous red galaxies (LRGs), and Quasar samples \cite{eBOSS:2020fvk,eBOSS:2020abk,eBOSS:2020lta,eBOSS:2020gbb}

    \item \textbf{Cosmic Chronometers (CC):} We also include the measures of $H(z)$ using the CC covariance matrix \cite{M_Moresco_2012, Moresco:2015cya, Moresco:2016mzx} which measures the Hubble evolution in an independent way.
    \item \textbf{PantheonPlus (PP) $\&$ SH0ES:} We include SNe-Ia measurements from the Pantheon+ compilation which contains 1701 light curves of 1550 distinct SNe-Ia in the redshift range $z \in [0.001, 2.26]$ \cite{Scolnic:2021amr, Brout:2022vxf}, referred to as PantheonPlus (PP). This sample also includes SH0ES Cepheid-calibrated host galaxy distance \cite{Riess:2021jrx}, denoted as R21. Thus, for the measurement of $H_0$, the SH0ES Cepheid host distance covariance matrix along with PP, we use the modified likelihood (refer Eq. 15 of \cite{Brout:2022vxf}), enabling us to place constraints on $H_0$.
    \item \textbf{MASERS:} We add the measurement of $H_0$ from Masers galaxy samples, UGC 3789, NGC 5765b, and NGC 4258 in the Hubble flow redshifts $z = 0.0116$, $0.0340$, and $0.0277$ respectively \cite{Reid_2009, Braatz_2010, Reid_2013, Kuo_2013, Gao_2016}. 
    \item \textbf{Growth rate data ($f\sigma_8$):} In addition to geometric probes, to investigate the matter density perturbations, we use growth rate data ($f\sigma_8$) measured from Redshift-space distortions (RSDs), as collected in \cite{Basilakos:2013nfa}.
    \item \textbf{Strong Gravitational Lensing Time Delay (SLTD)} We also add observations of the Hubble constant $H_0$ from time-delay gravitational lensing as measured by H0LiCOW experiments \cite{Bonvin:2016crt}.
\end{itemize}
To constrain the parameter space of the interaction with curvature (Coupled+$\Omega_K$) cosmological scenario, we implement a modified version of the Markov Chain Monte Carlo (MCMC) code, emcee: the MCMC Hammer \cite{Foreman-Mackey_2013} package. Table \ref{table:flat priors} lists the priors imposed on the cosmological parameters. During the analysis, the Hubble constant is assumed to be $H_0 = 100 h$ km s$^{-1}$ Mpc$^{-1}$, where $h$ is a dimensionless parameter. The baryon density parameter, $\Omega_{b_0}$, and the spectral index, $n_s$ are fixed at 0.045 and 0.96, respectively, according to \textit{Planck} (2018) \cite{Planck:2018vyg}.
It should be noted that we have employed positive priors on the interaction parameter. However, the results remain unchanged even if we opt for negative or larger priors instead.

\begin{table}[!ht]
\small
\centering
\addtolength{\tabcolsep}{8pt} 
\renewcommand{\arraystretch}{1.2} 
\begin{tabular}{|c | c|}
 \toprule
 \textbf{\textit{Parameters}}  & \textbf{\textit{Priors}} \\  
 \midrule
 $\Omega_{\phi_i}*10^{-9}$ & [0.05, 4.0]  \\ 
 $\Omega_{\text{r}_i}$ & [0.09, 0.19]  \\ 
 $\lambda_i$ & [$10^{-4}$, 3.0]  \\
 $m_i$ & [$10^{-6}$, 0.08] \\    
 $h$ & [0.4, 1.0]  \\
 $r_{\text{drag}}$ & [120, 180] Mpc \\
 $\sigma_{8}$ & [0.5, 1.0]  \\  
 $\Omega_K$ & [-1.0, 1.0] \\
 $M$ & [-21.0, -18.0] \\
\bottomrule
\end{tabular}
\caption{ Flat priors on the cosmological parameters
 used in this analysis.}
\label{table:flat priors}
\end{table}

\section{Results and Discussion}\label{results and discussion}
\begin{figure}[ht!]
    \centering
    \includegraphics[width=1.0\textwidth]{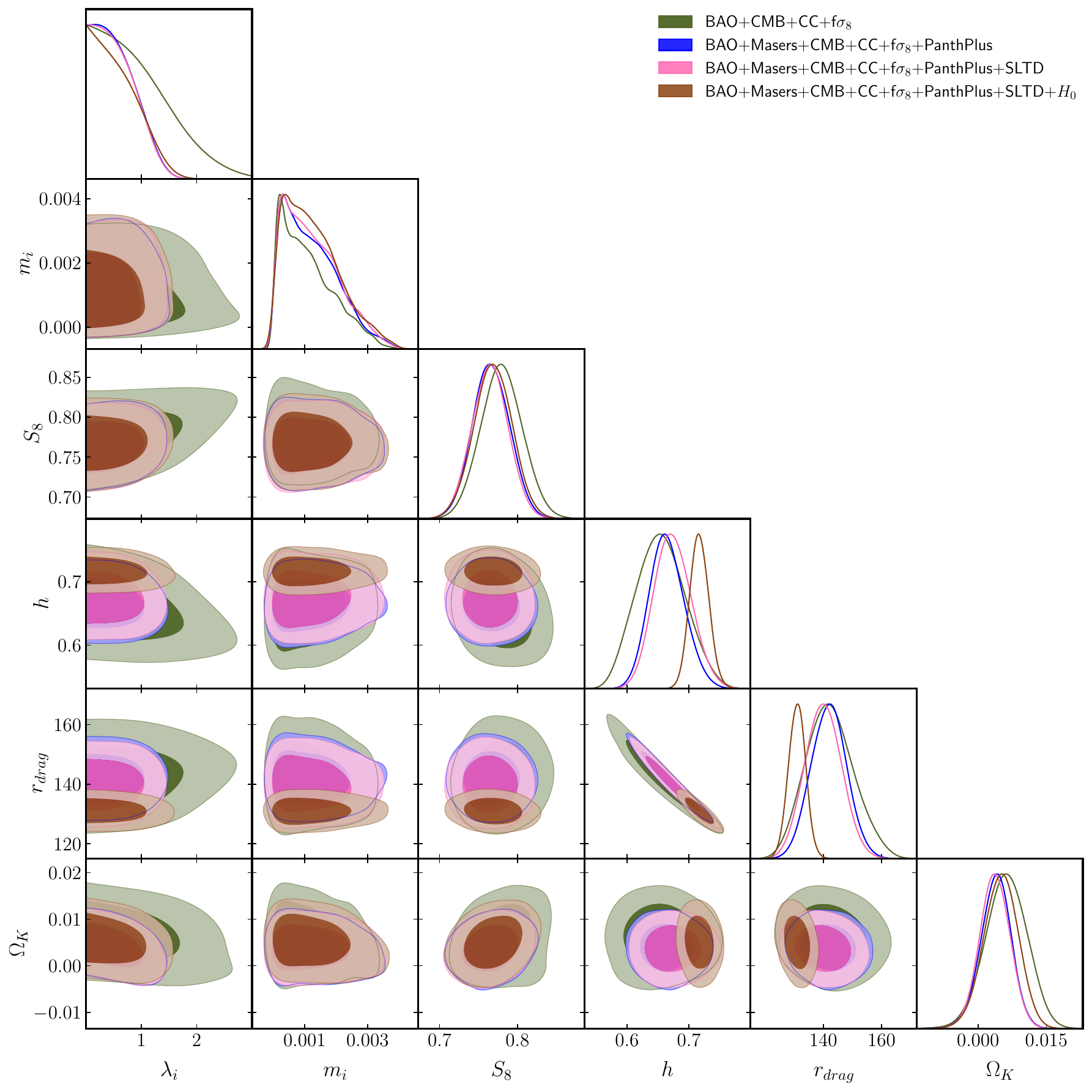}

    \caption{Contour plots at 68$\%$ and 95$\%$ CL and corresponding 1D marginalized posterior distributions for the most important parameters obtained from the MCMC analysis within the present \textbf{Coupled+$\Omega_k$} cosmological scenario across various data combinations.}
    \label{fig:merging_coupled}
\end{figure}

We present in Table \Ref{mean with error_curved} the results obtained within the coupled and curved cosmology (Coupled+$\Omega_k$) for four observational data combinations. We present the mean and 1$\sigma$ bounds (68$\%$ confidence levels (CL)) on the different cosmological parameters for the \textbf{\textit{BASE}} combination, representing `\textit{BAO+CMB+ CC+$f\sigma_8$}'. \textbf{\textit{BASE}} combination represent our baseline data. We subsequently add `\textit{Masers} and \textit{PantheonPlus (PP)}', `\textit{SLTD}', and `\textit{$H_0$}' data to the \BASE combination to study their effects on parameter constraints. The corresponding contour plots and 1D marginalized posterior distribution for different cosmological parameters are shown in Fig. \ref{fig:merging_coupled}. 

\subsection{Model Parameters in Coupled+$\Omega_k$}\label{Model Parameter and physical implications} 
The parameter $\lambda_i$ indicate how steeply the potential associated with the dark energy field, $\phi$ changes. The effect of variations in $\lambda_i$ can be seen on dark energy equation of state (DE EoS) parameter $\omega_\phi$, representing dark energy evolution, and cosmic structure formation and growth parameter, $S_8$, as shown in Table \ref{mean with error_curved}. Higher $\lambda_i$ value \big(0.90$^{+0.35}_{-0.82}$\big) for \textit{BASE} data combination indicates corresponding higher $\omega_\phi$ value. Whereas lower $\lambda_i$ \big(0.55$^{+0.22}_{-0.50}$ \big), for `\textit{BASE+Masers+PP+SLTD}' combination indicates lower $\omega_\phi$ value (more negative) for the same data combination. Similar effects are observed on structure growth parameter $S_8$, indicating higher $S_8$ for higher $\lambda_i$ and lower $S_8$ for low $\lambda_i$ values for the same data combination. 
We also found that the coupling parameter $m_i$, peaks at the low value, for all the four data combinations, as shown in Table \ref{mean with error_curved}. This small coupling magnitude indicates preference for a significantly weak interaction between dark energy and dark matter by the cosmological datasets in the Coupled+$\Omega_K$ case. However, these findings are model-dependent and may vary for other coupling models.
\begin{table}[!ht]
\small
\centering
\addtolength{\tabcolsep}{-5.0pt} 
\renewcommand{\arraystretch}{1.6}
\begin{tabular}{ c c c c c c } 
\toprule
\textbf{\textit{Parameters}} & \textbf{\textit{BASE}} & \multirow{1}{9em}{\centering \textbf{\textit{+Masers}} \\ \textbf{\textit{+PantheonPlus (PP)}}} & \multirow{1}{8em}{\centering \textbf{\textit{+Masers+PP}} \\ \textbf{\textit{+SLTD}}} & \multirow{1}{8em}{\centering \textbf{\textit{+Masers+PP}} \\ \textbf{\textit{+SLTD+$H_0$}}} \vspace{0.2cm} \\ 
\midrule
$\Omega_{\phi_i}*10^{-9}$ &  0.89$\pm$0.25 & 0.92$^{+0.29}_{-0.23}$ &  0.92$^{+0.30}_{-0.24} $  & 0.90$^{+0.29}_{-0.24}$ \\
$\Omega_{r_i}$ & 0.157$\pm$0.014 & 0.153$^{+0.011}_{-0.010}$ & 0.150$^{+0.012}_{-0.010}$  & 0.134$\pm$0.006 \\
$\lambda_i$ & 0.90$^{+0.35}_{-0.82}$ & 0.55$^{+0.21}_{-0.51}$ & 0.55$^{+0.22}_{-0.50}$  & 0.58$^{+0.22}_{-0.54}$ \\
$m_i$ & 0.0011$^{+0.00039}_{-0.00110}$  & 0.0012$^{+0.00047}_{-0.00110}$&  0.0012$^{+0.00049}_{-0.00110}$ &  0.0013$^{+0.00050}_{-0.00110}$ \\
$S_8$ &  0.779$^{+0.027}_{-0.026}$ & 0.766$^{+0.023}_{-0.024}$ & 0.764$\pm$0.023 & 0.768$\pm$0.024 \\ 
$\Omega_{dm}$ & 0.341$^{+0.0141}_{-0.0143}$  & 0.328$^{+0.0086}_{-0.0085}$ & 0.327$^{+0.0087}_{-0.0088}$ & 0.330$^{+0.0104}_{-0.0105}$ \\ 
$\omega_{\phi}$ & -0.931$^{+0.065}_{-0.066}$  & -0.971$^{+0.027}_{-0.030}$ & -0.971$^{+0.026}_{-0.029}$ & -0.969$^{+0.029}_{-0.032}$ \\ 
$h$ & 0.655$^{+0.038}_{-0.042}$ & 0.665$^{+0.026}_{-0.030}$  & 0.674$^{+0.028}_{-0.032}$ & 0.717$\pm$0.015\\ 
\multirow{1}{*}{$r_{drag}$ (Mpc)} & 142.3$^{+7.5}_{-8.8}$ & 141.9$\pm$6.0 & 140.0$\pm$6.2 & 131.0$\pm$2.9 \\ 
$\Omega_{k}$ & 0.0062$\pm$0.0046 & 0.0040$^{+0.0035}_{-0.0031}$ & 0.0037$\pm$0.0033 & 0.0050$\pm$0.0038 \\ 
$\Delta \chi^2_{min}$ & 15.091   & 6.690 &  -1.269  & 0.321   \\
$\Delta AIC$ &  19.091  & 10.690  & 2.730  &  4.321 \\
\bottomrule
\end{tabular}
\caption{\label{mean with error_curved} Mean values with 1$\sigma$ bound (68$\%$ CL) on the parameters within the \textbf{Coupled+$\Omega_K$} scenario from various datasets. The $\Delta \chi^2_{min}$ and $\Delta AIC$ are calculated with respect to the $\Lambda$CDM+$\Omega_K$ model for the very same dataset. The negative values indicate the preference for the Coupled+$\Omega_K$ case, while the positive values for the $\Lambda$CDM+$\Omega_K$ case. Here, `\textbf{\textit{BASE}}' represents `\textit{BAO+CMB+CC+$f\sigma_8$}' dataset.} 
\end{table}
\subsection{The $H_0$ Tension}
In the curved cosmology analysis presented in Table \ref{mean with error_curved}, we have found, that `\textit{BAO+CMB+CC+$f\sigma_8$}’ (\textit{BASE}) constrains $H_0$ to 65.5$^{+3.8}_{-4.2}$ km/s/Mpc at 68$\%$ CL, which is 0.4$\sigma$ away from \textit{Planck} 2018 + $\Lambda$CDM results ($H_0$ = 67.37 $\pm$0.54 km/s/Mpc) \cite{Planck:2018vyg} (hereafter `\textit{Planck} 2018 + $\Lambda$CDM results' will be referred to as `P18+$\Lambda$CDM' collectively). 
However, the same $H_0$ constraints indicate a 1.5$\sigma$ difference with the SH0ES determination of $H_0$ ($H_0$ = 73.04 $\pm$ 1.01 km/s/Mpc) \cite{Riess:2021jrx} (hereafter `SH0ES determination of $H_0$' will be referred to as `R21'). 
This improvement in the Hubble constant, $H_0$ value, which shifts $H_0$ towards the R21 value within 1.5$\sigma$, is primarily due to an increased volume in the parameter space rather than an actual rise in the mean value of $H_0$. With each addition to the \textit{BASE} dataset, the constraints strengthen and the mean value shifts towards the R21 value, maintaining the tension with R21 at the level of $\sim$1.7$\sigma$. When the SH0ES constraint is included (last column of Table \ref{mean with error_curved}), the fit is increased to $H_0$ = 71.7$\pm$1.5 km/s/Mpc, indicating a less than 2.2$\sigma$ deviation from P18+$\Lambda$CDM and a 0.6$\sigma$ shift towards R21. These findings are also evident in Fig. \ref{fig:merging_coupled}. We noticed that the $H_0$ results obtained from fitting the `Coupled+$\Omega_k$' model using the dataset `\textit{BASE+Masers+PP+SLTD+$H_0$}' demonstrate slight reductions in 
difference with both P18+$\Lambda$CDM \cite{Planck:2018vyg} and R21 \cite{Riess:2021jrx} analyses compared to the results found in the flat case interacting model discussed in \cite{Patil:2023rqy}.




\subsection{The $S_8$ Tension}\label{the S8 tension} 
In the $S_8$ analysis, the \textit{BASE} estimates $S_8$ to 0.779$^{+0.027}_{-0.026}$, with 1.4$\sigma$ away from P18+$\Lambda$CDM \big($S_8$ = $0.830\pm0.013$\big) \cite{Planck:2018vyg}. However, this shift in $S_8$ measurement is less than 0.5$\sigma$ towards lensing observation estimates \cite{Heymans:2020gsg,DES:2021wwk}.  
The Masers and PantheonPlus addition to \textit{BASE} further provides unprecedented precision on $S_8$, preferring lower $S_8$ \big($S_8$ = 0.766$^{+0.023}_{-0.024}$\big), with a shift towards DES-Y3 estimates \cite{DES:2021wwk} \big($S_8$ = 0.776$\pm$0.017\big) by less than 0.3$\sigma$ and deviation from P18+$\Lambda$CDM by 1.8$\sigma$.  We noticed that for this combination, the $S_8$ estimates closely aligns with the value from KiDS-1000-BOSS \big($S_8$ = 0.766$^{+0.020}_{-0.014}$\big) \cite{Heymans:2020gsg}. The inclusion of SLTD (third column of Table \ref{mean with error_curved}) predicts lower $S_8$ value than any other dataset shown in Table \ref{mean with error_curved}. By considering the combined analysis `\textit{BASE+Masers+PP+SLTD+$H_0$}' (all data), we find $S_8$ = 0.768$\pm$0.024. This value indicates a reduced difference within 2$\sigma$ range from P18+$\Lambda$CDM \cite{Planck:2018vyg}, compared to the 2.5$\sigma$ discussed in the flat case interacting model \cite{Patil:2023rqy}. Notably, this $S_8$ value is very close to those from DES-Y3 \cite{DES:2021wwk} (within 0.2$\sigma$) and KiDS-1000-BOSS \cite{Heymans:2020gsg} (within 0.1$\sigma$), improving upon the 0.5$\sigma$ and 0.3$\sigma$ differences observed in the flat case interacting model \cite{Patil:2023rqy}. Note that even though we have not used the KiDS-1000-BOSS and DES-Y3 survey results while performing the Coupled+$\Omega_K$ model analysis, the $S_8$ derived with the Coupled+$\Omega_K$ framework are qualitatively same as the one obtained by using the KiDS-1000-BOSS and DES-Y3 data in their models \cite{Heymans:2020gsg,DES:2021wwk}.

\subsection{Parameter Correlations in Coupled+$\Omega_K$}\label{Parameter Correlations in coupled+OmegaK}
\begin{figure}
    \centering
    \begin{minipage}[t]{0.46\textwidth}
        \centering
        \includegraphics[width=1.03\textwidth]{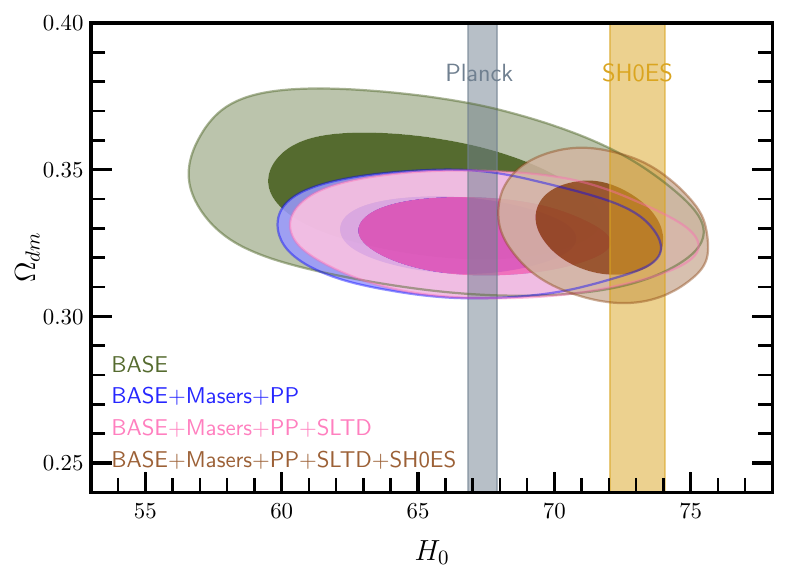} 
    \end{minipage}\hfill
    \begin{minipage}[t]{0.46\textwidth}
        \centering
        \includegraphics[width=1.02\textwidth]{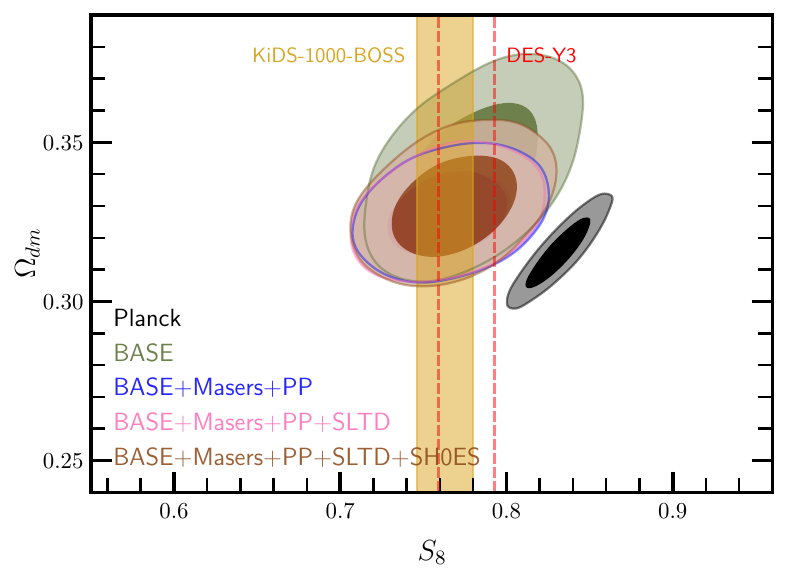} 
    \end{minipage}
    \caption{\textbf{Left:} 68$\%$ and 95$\%$ CL constraints on $H_0$-$\Omega_{dm}$ parameter space from various datasets shown in the figure for Coupled+$\Omega_K$ scenario. The `Gray' and `Golden' bands represent the 68$\%$ CL constraints on $H_0$ from \Planck and SH0ES analyses, respectively. \textbf{Right:} 68$\%$ and 95$\%$ CL constraints on $S_8$-$\Omega_{dm}$ parameter space from various datasets shown in the figure for Coupled+$\Omega_K$ scenario. The `Golden' band and `Red-dashed' line represent the 68$\%$ CL constraints on $S_8$ from KiDS-1000-BOSS and DES-Y3 analyses, respectively. The `Black' contours show 68$\%$ and 95$\%$ CL on $S_8$ from \textit{Planck} 2018.}
    \label{fig:h0-s8_andDEEoScurvedspacetime}
\end{figure}

We also observed the change in $\Omega_K$ for each dataset. These changes in $\Omega_K$ are associated with the relative shifts in $S_8$ and $\Omega_{dm}$ (as shown in Table \ref{mean with error_curved}), indicating a slightly higher value of $\Omega_{dm}$ = 0.341$^{+0.0141}_{-0.0143}$ and $S_8$ = 0.779 $^{+0.027}_{-0.026}$ corresponding to higher $\Omega_K$ \big($\Omega_K$ = 0.0062$\pm$0.0046\big) for the \textit{BASE} data combination and a shift towards lower value of $\Omega_{dm}$ = 0.327$^{+0.0087}_{-0.0088}$ and $S_8$ = 0.764$\pm$0.023 corresponding to lower $\Omega_K$ \big($\Omega_K$ = 0.0037$\pm$0.0033\big) for `\textit{BASE+Masers+PP+SLTD}' combination, with more precise constraints. The $S_8$ and $\Omega_{dm}$ correlation results are presented in the right panel of Fig. \ref{fig:h0-s8_andDEEoScurvedspacetime}. It is to be noted that the $S_8$ results presented in subsection \ref{the S8 tension} show a difference of $\sim$1.4$\sigma$ for the \textit{BASE} dataset and $\sim$2$\sigma$ for the `\textit{BASE+Masers+PP+SLTD+$H_0$}' dataset from the P18+$\Lambda$CDM. However, when examining the $S_8$-$\Omega_{dm}$ contour plots (Fig. \ref{fig:h0-s8_andDEEoScurvedspacetime}, right panel), the difference in those contours exceeds 2$\sigma$ from the $Planck$ CMB contour. This increase in the difference appeared from the correlation between $S_8$-$\Omega_{dm}$. Furthermore, due to the anti-correlation between $H_0$ and the dark matter density parameter $\Omega_{dm}$, we noticed the $\Omega_{dm}$ transition from higher to lower values correspond to lower to higher values of $H_0$, excepts for the combined analysis `\textit{BASE+Masers+PP+SLTD+$H_0$}', as shown in Table \ref{mean with error_curved}. In this case, the lack of correlation between $H_0$ and $\Omega_{dm}$ indicates their independent evolution, effectively breaking the degeneracy for the combined analysis (the last column of Table \ref{mean with error_curved}). These results are presented in the left panel of Fig. \ref{fig:h0-s8_andDEEoScurvedspacetime}.


Similarly, the lack of correlation is also observed in $H_0$ and $\Omega_K$ parameter space for the `\textit{BASE+ Masers+PP+SLTD+$H_0$}' dataset. Additionally, we noticed the high $H_0$ value correspond to positive $\Omega_K$ for the combined analysis `\textit{BASE+Masers+PP+SLTD+$H_0$}', similar to the trend observed in the $\Lambda$CDM+$\Omega_K$ model (refer Table \ref{mean with error_allmodel}). This indicates that, despite the added complexity introduced by the coupling and curvature parameter, the relationship between $H_0$ and $\Omega_K$ remains consistent, favoring a slightly positive curvature parameter for higher $H_0$, as shown in the last column of Table \ref{mean with error_curved}.

\subsection{The $\Omega_K$ and the Impact on $\omega_{\phi}$}\label{The Omega_K and the Impact on omega_phi}
Now, we investigate the curvature parameter, $\Omega_K$ and how it affects the dark energy equation of state parameter, $\omega_{\phi}$ constraints. In the `Coupled+$\Omega_K$' model, for `\textit{BASE}' dataset we get an evidence of an open Universe with the constraint on the curvature parameter reading $\Omega_{k}$ = 0.0062$\pm$0.0046 and a non-zero coupling strength at 68 $\%$ CL. In this case, the corresponding $\omega_{\phi}$ is found to be $\omega_{\phi}$ = -0.931$^{+0.065}_{-0.066}$, which is moderately higher than the $\omega_{\phi}$ obtained in other datasets presented in Table \ref{mean with error_curved}. The preference for the curved spacetime is found to be consistent at 68$\%$ CL for `\textit{BASE+Masers+PP}', `\textit{BASE+Masers+PP+SLTD}', and `\textit{BASE+Masers+PP+SLTD+$H_0$}' combinations, with only a mild indication for
the presence of coupling in the dark sector. As $\Omega_{k}$ prefers lower values, we noticed more negative values of $\omega_{\phi}$ for all these data combinations compared to the \textit{BASE} combination. Also, the constraint on $\Omega_k$ from the `\textit{BASE+Masers+PP+SLTD}' \big($\Omega_k = 0.0037\pm0.0033$\big) is more precise than those obtained using other datasets. This improved precision is reflected in the tighter constraints on $\omega_\phi$ \big($\omega_\phi$ = -0.971$^{+0.026}_{-0.029}$\big), shown in the fourth column of Table \ref{mean with error_curved}, which is in perfect agreement with the cosmological constant case ($\omega_{\phi} = -1$) at 68$\%$ CL. 

\begin{figure}[ht!]
    \centering
    \includegraphics[width=0.6\textwidth]{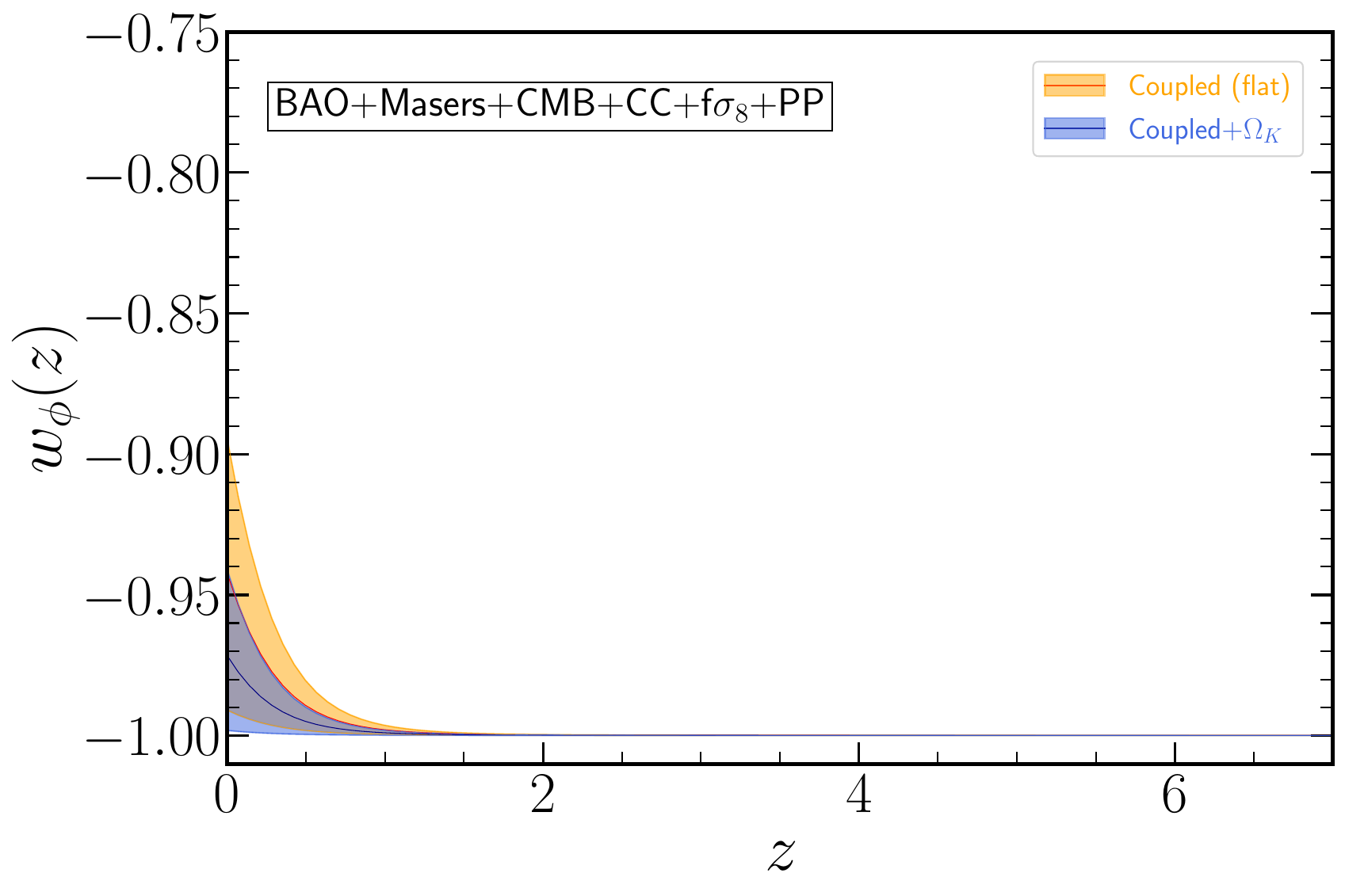}
    \caption{Evolution of the dark energy equation of state parameter $\omega_{\phi}(z)$ with respect to redshift for (flat) coupled and Coupled+$\Omega_K$ scenarios. The orange and blue regions represent the 68$\%$ CL on $\omega_{\phi}(z)$ for the flat coupled and Coupled+$\Omega_K$ cases, respectively, using the \textit{`BAO+Masers+CMB+CC+$f\sigma_8$+PP'} combination.}
    \label{fig:wphi_flatandnonflat}
\end{figure}

When analyzed, we found that the $\omega_{\phi}$ values, in the `Coupled + $\Omega_K$' scenario, are lower, compared to those obtained in the flat interacting scenario in \cite{Patil:2023rqy}. However, the estimates of $\omega_{\phi}$ derived from both the `Coupled + $\Omega_K$' scenario and the flat interacting scenario are within $\sim$ 1$\sigma$ bounds of each other. We have also shown the comparison of our result with the previous works on the constraints of $\omega_{\phi}$ in Fig. \ref{fig:wphi_flatandnonflat} for \textit{`BAO+Masers+CMB+CC+$f\sigma_8$+PP'} dataset, where $\omega_{\phi}$ is found to be \big($\omega_\phi$ = -0.9422$^{+0.0485}_{-0.0475}$\big) in flat interacting model.

\subsection{Model Probability Assessment}
Finally, we discuss the comparison between the Coupled+$\Omega_K$ model and the $\Lambda$CDM+$\Omega_K$ model in Table \Ref{mean with error_curved}. We compare the minimum $\chi^2$ and Akaike information criterion (AIC) values obtained for each model across the respective datasets, where $AIC = \chi^2_{\text{min}} + 2 \times d$, and $d$ is the number of parameters in the model. The lowest chi-squared and AIC values indicate the best-fitting model, closely matching the measured data. In Table \Ref{mean with error_curved}, negative values of $\Delta \chi^2_{min}$ and $\Delta AIC$ indicate a preference for the Coupled+$\Omega_K$ model, while positive values indicate a preference for the $\Lambda$CDM+$\Omega_K$ model. 
With a positive $\Delta AIC$, we observed higher values of $AIC$ for the Coupled+$\Omega_K$ model compared to the $\Lambda$CDM+$\Omega_K$ framework. This suggests that the Coupled+$\Omega_K$ model provides a weaker fit to the observational data across all dataset combinations, and is therefore disfavored by the $AIC$ approach. Furthermore, in the \textit{`BASE+Masers+PP+SLTD'} case, $\Delta \chi^2_{\text{min}} = -1.269$ indicates preference for Coupled+$\Omega_K$ model over $\Lambda$CDM+$\Omega_K$. However, based on the value of $\Delta AIC = 2.730$ for this dataset, the $\Lambda$CDM+$\Omega_K$ model exhibits better concordance with the data, indicating no preference for Coupled+$\Omega_K$ one. These inconclusive results are insufficient to definitively favor one model over the other, particularly for the `\textit{BASE+Masers+PP+SLTD}' dataset.



\subsection{Cosmological Constraints from Different Models }
We now compare the constraints on the cosmological parameters obtained from five different models: flat $\Lambda$CDM (denoted by $\Lambda$CDM (Flat)), $\Lambda$CDM+$\Omega_k$, Un-coupled+$\Omega_k$, Coupled+$\Omega_k$, and flat coupled (denoted by Coupled (Flat)) for the `\textit{BASE+Masers+PP+SLTD+$H_0$}' combination, as shown in Table \ref{mean with error_allmodel}. Here, $\Lambda$CDM+$\Omega_k$ represents the $\Lambda$CDM framework in a curved universe, and Un-coupled+$\Omega_k$ represents a scenario without coupling (interaction), where the coupling parameter $m_i$ in the Coupled+$\Omega_k$ model is set to zero.

\begin{table}[!ht]
\small
\centering
\addtolength{\tabcolsep}{-2.0pt} 
\renewcommand{\arraystretch}{1.6}
\begin{tabular}{ c c c c c c c } 
\toprule
\textbf{\textit{Parameters}}  &  $\Lambda$CDM (Flat) & $\Lambda$CDM+$\Omega_k$ & Un-coupled+$\Omega_k$ & Coupled+$\Omega_k$  & Coupled (Flat)\\ 
\midrule
$S_8$ & 0.754$\pm$0.022  & 0.774$\pm$0.024 & 0.758$\pm$0.024& 0.768$\pm$0.024 & 0.761$\pm$0.023\\ 
$\Omega_{dm}$ & 0.318$\pm$0.0073 &  0.335$\pm$0.0110 & 0.331$^{+0.0104}_{-0.0103}$ & 0.330$^{+0.0104}_{-0.0105}$ & 0.322$^{+0.0081}_{-0.0082}$  \\ 
$h$ & 0.722$\pm$0.013 & 0.717$\pm$0.013  & 0.718$\pm$0.015 & 0.717$\pm$0.015 & 0.718$\pm$0.016 \\ 
\multirow{1}{*}{$r_{drag}$ (Mpc)} & 132.1$\pm$2.5 & 131.6$\pm$2.6  & 130.7$\pm$3.0 & 131.0$\pm$2.9 & 132.0$\pm$3.0 \\ 
$\Omega_{k}$ & - & 0.0510$\pm$0.0180  & 0.0056$\pm$0.0037 & 0.0050$\pm$ 0.0038 & -\\ 
$m_i$ & - & - & -&  0.0013$^{+0.00050}_{-0.00110}$ & 0.0014$^{+0.00060}_{-0.00120}$ \\ 
$\Delta \chi^2_{min}$ & 0  & -4.869  & 28.351  & -4.548 & 25.099  \\
$\Delta AIC$ & 0  & -2.869  & 32.351  & 1.451 & 29.099 \\
\bottomrule
\end{tabular}
\caption{\label{mean with error_allmodel} Mean values with 1$\sigma$ bound obtained on the most relevant parameters within the $\Lambda$CDM (Flat), $\Lambda$CDM+$\Omega_k$, Un-coupled+$\Omega_k$, Coupled+$\Omega_k$, and Coupled (Flat) scenarios for `\textbf{\textit{BASE+Masers+PP+SLTD+$H_0$}}' dataset. The $\Delta \chi^2_{min}$ and $\Delta AIC$ are calculated with respect to the $\Lambda$CDM model. The sign convention is same as in Table \ref{mean with error_curved}.} 
\end{table}

In our analysis, we observed that the $H_0$ estimates in the flat $\Lambda$CDM and $\Lambda$CDM+$\Omega_K$ models are approximately 1.2 times more precise than those in the other three modeling scenarios, as shown in Table \ref{mean with error_allmodel}. However, these estimates exhibit a $\sim$2.7$\sigma$ difference from the P18+$\Lambda$CDM result \cite{Planck:2018vyg}. In contrast, the Un-coupled+$\Omega_K$, Coupled+$\Omega_K$, and Coupled (Flat) models show a $\sim$2.2$\sigma$ difference with P18+$\Lambda$CDM. The constraints obtained on $H_0$ align more closely with the R21 result \cite{Riess:2021jrx} and remain consistent across all models within 0.6$\sigma$, except for the $\Lambda$CDM+$\Omega_k$ model, where $H_0 = 71.7\pm1.3$ km/s/Mpc indicates a 1$\sigma$ deviation from the R21 value. Please note that the derived $H_0$ is higher for all models in Table \ref{mean with error_allmodel}, so the Coupled+$\Omega_k$ model is not an exception. Similarly, the constraints on $S_8$ in the flat $\Lambda$CDM model exhibit better precision (a lower and a well-constrained $S_8$ estimation) than those in the other four models, as shown in Table \Ref{mean with error_allmodel}. In this case, $S_8$ has a lower value (0.754$\pm$ 0.022 at 68$\%$ CL), shifting it 2.2$\sigma$ away from P18+$\Lambda$CDM \cite{Planck:2018vyg}, $\sim$0.7$\sigma$ closer to DES-Y3 \cite{DES:2021wwk}, and $\sim$0.3$\sigma$ closer to KiDS-1000-BOSS \cite{Heymans:2020gsg}. Consequently, the $\Lambda$CDM+$\Omega_k$ bounds on $S_8$ are 0.774$\pm$0.024 at 68$\%$ CL, showing a slight decrease in the difference (1.6$\sigma$) with P18+$\Lambda$CDM compared to the flat $\Lambda$CDM case (as discussed previously). Additionally, this analysis shows close alignment with DES-Y3 (0.1$\sigma$) while $S_8$ retaining its consistency of $\sim$0.3$\sigma$ with KiDS-1000-BOSS survey as in the flat $\Lambda$CDM model\footnote{Note that the slight difference in parameter constraints in Coupled (Flat) model, in Table \ref{mean with error_allmodel} are due to the addition of new datasets, SLTD and BAO SDSS DR16, which were not considered in our previous study \cite{Patil:2023rqy} for the flat case interacting model.}. 



We also observed that the curvature density parameter, $\Omega_K$, indicates a preference for an open universe in the $\Lambda$CDM+$\Omega_K$ case which is found to be $\Omega_K$ = 0.0510$\pm$ 0.0180 at 68$\%$ CL (third column of Table \ref{mean with error_allmodel}). This preference for a curved spacetime persists in Un-coupled+$\Omega_K$ and Coupled+$\Omega_K$ scenarios, with significant reduction in $\Omega_K$ values (both mean and error bars) compared to $\Lambda$CDM+$\Omega_K$. These precise constraints narrow down the $\Omega_K$ parameter space. The effects of curvature are also witnessed on the parameters $S_8$ and $\Omega_{dm}$ in Table \ref{mean with error_allmodel}. In the flat $\Lambda$CDM model, the absence of $\Omega_K$ indicates tightened constraints on $S_8$ and $\Omega_{dm}$, while the introduction of curvature broadens the parameter space for these parameters in $\Lambda$CDM+$\Omega_k$, Un-coupled+$\Omega_k$, and Coupled+$\Omega_k$ models. The DM density parameter reading $\Omega_{dm}$ = 0.322$^{+0.0081}_{-0.0082}$ at 68$\%$ CL in the flat coupled case, we observed moderate broadening in the 
parameter space compared to the flat $\Lambda$CDM model due to the effect of coupling complexities. However, this constraint is still better than that obtained in the other three scenarios where the curvature component is included. 

In this analysis, the model comparison statistics are computed relative to the flat $\Lambda$CDM using the `\textit{BASE+Masers+PP+SLTD+$H_0$}' dataset. We noticed that with $\Delta AIC$ = -2.869, the $\Lambda$CDM+$\Omega_K$ model fits the observations better than other models presented in Table \ref{mean with error_allmodel} and the same is also preferred over the flat $\Lambda$CDM. In addition, positive values of both, $\Delta \chi^2_{min}$ and $\Delta AIC$ in Un-coupled+$\Omega_k$ and Coupled (Flat) cases show that the data significantly prefer the flat $\Lambda$CDM picture over these models. We also noted that the $\Delta \chi^2_{min}$ in the Coupled+$\Omega_k$ model is negative. However, this does not indicate a preference for this model over the flat $\Lambda$CDM one, as can be inferred from the value of $\Delta AIC$ ($\Delta AIC > 0$) shown in the fifth column of Table \ref{mean with error_allmodel}. 






\section{Conclusions}\label{conclusions}
In this paper, we study an extension of the coupled dark energy-dark matter model having dynamic interaction $Q = F,_{\phi} \rho_{dm}\dot{\phi}$ by introducing curvature component in the universe. We aim to investigate the effects of curvature on the coupled quintessence model and its associated (free and derived) parameters, in particular $H_0$ and $S_8$.

We tested our model with several combinations of data and analyzed the influence of each dataset on the cosmological parameters. Though weak, we observed a coupling which is different from zero today. We noticed large error bars in computing the $H_0$ value from \textit{BASE} combination. These error bars are reduced with the addition of `\textit{Masers+PP}' data to the previous dataset. With further addition of `\textit{SLTD}' and SH0ES' `\textit{$H_0$}' data, the mean value is increased, shifting the consistency more towards the R21 analysis. The $S_8$ is well constrained for all the datasets considered in this analysis in Table \ref{mean with error_curved}. The addition of subsequent data shows $S_8$ consistency more toward the large-scale structure observations. We also mentioned the dark matter energy density $\Omega_{dm}$ parameter and presented its respective correlation with $H_0$ and $S_8$, in terms of $H_0$-$\Omega_{dm}$ and $S_8$-$\Omega_{dm}$ planes in Fig. \ref{fig:h0-s8_andDEEoScurvedspacetime}. Furthermore, the bounds on $\Omega_K$ preferred an open Universe at 68$\%$ CL under the \textit{BASE} dataset in Table \ref{mean with error_curved}, and this preference for the curved Universe remains the same for all other data considerations. 
Later, we presented the impact of the curvature component on the time evolution of the DE EoS parameter, $\omega_{\phi}$ in Sec. \ref{The Omega_K and the Impact on omega_phi} 
and shown the comparison between flat coupled and Coupled+$\Omega_K$ models in Fig. \ref{fig:wphi_flatandnonflat}. While measuring the goodness of fit of the Coupled+$\Omega_K$ model to the data with respect to the $\Lambda$CDM+$\Omega_K$ in Table \ref{mean with error_curved}, we have found no evidence in favor of Coupled+$\Omega_K$ by any of the data combinations. Here, the fit determined by AIC indicates a preference for the $\Lambda$CDM+$\Omega_K$ framework because $\Delta AIC > 0$ in all cases.


In the end, we presented and compared the cosmological constraints obtained from different model analyses in Table \ref{mean with error_allmodel}. In all models that we considered, we found that the Hubble tension parameter, $H_0$ indicates close alignments with R21 analysis, except 1$\sigma$ deviation in the $\Lambda$CDM+$\Omega_K$ model due to its ability of finding a lower and precise constraint on $H_0$ than any other models, as shown in Table \ref{mean with error_allmodel}. Also, we found that the matter structure growth parameter, $S_8$, in the $\Lambda$CDM+$\Omega_K$ model shows quite a negligible difference with DES-Y3 results, and in this case, the model is favored over $\Lambda$CDM \big($\Delta \chi^2_{min}=-4.869$ and $\Delta AIC=-2.869$\big). We found the similar constraints on $S_8$ (within 0.2$\sigma$ bounds with DES-Y3 and KiDS-1000-BOSS survey results) when analysing Coupled+$\Omega_K$ case, using the combined analysis `\textit{BASE+Masers+PP+SLTD+$H_0$}'. However, in this case, the model is not favored by the data because $\Delta AIC=1.451$, even though we obtained $\Delta \chi^2_{min}=-4.548$. 

Following the above analyses, we noticed that the coupled case, with the presence of non-zero curvature (Coupled+$\Omega_K$ model), can provide a compelling solution to the $H_0$ and $S_8$ tensions. Even from the model comparison perspective, we found that the coupled+$\Omega_K$ model, using all the dataset has a chance to better fit the observations \big($\Delta \chi^2_{min}=-4.548$\big) and can be a preferred model relative to the flat $\Lambda$CDM framework (Table \ref{mean with error_allmodel}). Besides this, considering the data analyses of model under assumption, our study demonstrates how the interaction between dark energy and dark matter within non-zero spatial curvature framework modifies the universe's dynamics. This includes the timing of dark energy dominance, its substantial effects on late-time accelerated expansion, and its role in structure formation and clustering, as reflected in parameters like $S_8$. These results reveal that observed cosmological tensions can be attributed to either modifications in the universe's geometry (spatial curvature) or interactions within the dark sector or both. Specifically, while the spatial curvature influences the cosmic energy budget and modifies the Hubble parameter and structure growth rate, interaction strength affects the equation of state for dark energy ($\omega_{de}$) and dark matter clustering. These effects, together, suggest that it may be better to relax the flatness assumption and the cosmological constant nature of dark energy when analyzing cosmological models, as they could provide solutions to a number of cosmological tensions. In this context, further in-depth research is needed at both theoretical and observational levels to establish the interaction model as a promising cosmological model, since such an interaction may lead to modifications in the clustering patterns of large-scale structures, potentially affecting clustering strength at various scales. To draw more robust conclusions, we need to improve upon the statistical errors, which can be achieved by including data from future high-quality observations such as from Euclid \cite{Amendola:2016saw} and DESI \cite{DESI:2022lza,DESI:2024mwx} surveys that promise enhanced precision for cosmological parameter measurements \cite{Perenon:2020jav,MNRAS.16585.x,Yankelevich:2018uaz}. In the future, we also plan to perform further analysis with other forms of interacting dark energy, such as phantom one, and use DESI BAO data which could further help alleviate these discrepancies.

\section*{Acknowledgments}
We thank the anonymous referee for his/her thoughtful remarks and suggestions which led to
material improvements in the manuscripts. We are grateful to Ruchika for her assistance during the computational part and for further discussion. This work is partially supported by DST (Govt. of India) Grant No. SERB/PHY/2021057. \vspace{0.3cm}\\
\textbf{Data availability:} No data is associated with the manuscript.

\section*{Appendix}

The following explains the Hubble evolution dependence on spatial curvature and dark sector coupling.\\
The measurement of the angular scale of the sound horizon ($\theta_s = r_s(z_{dec})/D_A(z_{obs})$) infers the Hubble expansion parameter, $H_0$, under the assumption of a given cosmological model. 
Here, $r_s(z_{dec})$ is the comoving sound horizon at the redshift $z_{dec}$ at which photons and baryons decouples, and $D_A(z_{obs})$ is the comoving angular diameter distance to the observation. $D_A(z_{obs})$ depends on the redshift-dependent expansion rate $H(z)$, 
as $D_A(z_{obs}) = \frac{1}{H_0} \int_0^{z_{obs}} \frac{dz}{E(z)}$, where $E(z)$ incorporates contributions from matter, curvature, and dark energy components, i.e. $E(z) = \sqrt{\Omega_{m}(1+z)^3+\Omega_{K}(1+z)^2+\Omega_{de}(z)}$ $ = \frac{H(z)}{H_0}$. To achieve a high $H_0$ value $(\sim 73Km/s/Mpc)$ consistent with local measurements \cite{Riess:2021jrx}, the late-universe solution ($z \leq 2$) requires modification involving new physics. While dynamical dark energy or interaction among dark sector components can alter the Hubble expansion rate, incorporating an ingredient in the form of spatial curvature offers an additional degree of freedom. This allows $H_0$ to approach a higher value today while preserving the integrated distance $D_A$. This may help to understand and find solution to the $H_0$ tension. Additionally, the spatial curvature can also alter the growth of density perturbations by modifying overall expansion rate, leading to reduce the discrepancy in the growth parameter $S_8$ between high-redshift (CMB) \cite{Planck:2018vyg} and low-redshift (weak lensing, galaxy clustering) \cite{Heymans:2020gsg,DES:2021vln,DES:2022ccp} observations.

\singlespacing
\bibliographystyle{ieeetr}
\bibliography{refs}

\end{document}